\documentstyle[aps,12pt,epsfig]{revtex}

\def\slash#1{#1 \hskip -0.5em / }
\newcommand{\fr}[1]{
             \frac{#1}}
\newcommand{\eq}{\equiv}

\newcommand{\chibar}{\overline{\chi}}

\newcommand{\ket}{{\cal i}}
\newcommand{\bra}{{\cal h}}
\newcommand{\be}{\begin{equation}}
\newcommand{\ee}{\end{equation}}
\newcommand{\bea}{\begin{eqnarray}}
\newcommand{\eea}{\end{eqnarray}}
\newcommand{\p}{\, +\,}

\newcommand{\e}{\, =\,}
\newcommand\pubnumberk{CERN-TH/2001-250 \\ OSLO-TP-6-01 \\
HIP-2001-49/TH\\
hep-ph/0109xxx}

\def\csumb{$^a$ CERN, Theory Division, CH-1211 Geneva 23, Switzerland\\
$^b$ Department of Physics, POB 1048, N-0316 Oslo, University of Oslo, Norway\\
$^c$Physics Department, POB 64, FIN-00014, University of Helsinki,
Finland}
\newcommand\pubblock{\rightline{\begin{tabular}{l} 
                                                   \pubnumberk\end{tabular}}}
\def\Title#1{\begin{center} {\Large\bf #1 } \end{center}}
\def\Author#1{\begin{center}{ \sc #1} \end{center}}
\def\Address#1{\begin{center}{ \it #1} \end{center}}

\newenvironment{Abstract}{\begin{quotation}  }{\end{quotation}}

\begin{document}
\begin{titlepage}
\pubblock
\vfill
\Title{A gluonic mechanism for $B \rightarrow D \eta'$}

\Author{J. O. Eeg$^{a,b}$, A. Hiorth$^{b}$ and A. D. Polosa$^c$}
\Address{\csumb}
\vfill
\begin{Abstract}
We present a calculation of the process $\bar{B}^0\rightarrow
D^0\eta^\prime$ within a heavy--light chiral quark model. We assume that
the $\eta^\prime$  has a large gluonic component, and its
coupling is described via the glue--glue--$\eta^\prime$ effective
vertex. The main contribution comes from the non-factorizable part
of the effective weak Lagrangian at quark level. We find,
within our model-dependent assumptions,  a branching ratio 
$B_r(\bar{B^0}\rightarrow D^0\eta^\prime)\, 
= \,$ (1.7 -- 3.3) $\times 10^{-4}$,
somewhat below the experimental upper bound $9.4 \times 10^{-4}$. 
\end{Abstract}
\vfill
\end{titlepage}
\eject \baselineskip=0.3in
\section{Introduction}

After the CLEO \cite{cleo1} results reporting a very large
branching ratio for the inclusive production of $\eta^\prime$ in
$B\to \eta^\prime X_s$ decays, Atwood and Soni \cite{atwood}
proposed an interesting explanation of the data. The suggested
mechanism is based on the subprocess $b\to sg^*\to s\eta^\prime g$,
where the virtual gluon $g^*$ emerging from the standard model
penguin couples to $\eta^\prime$ via an effective $g g^*
\eta^\prime$ vertex related to the gluonic triangle anomaly.
 The structure of this vertex was re-examined in
\cite{hou}, where the running of the effective coupling of
$\eta^\prime$ to gluons, assumed to be constant in \cite{atwood},
is also taken into account. The possibility that the $g g^*
\eta^\prime$ vertex could dangerously be affected by 
non-controllable non-perturbative effects was soon after discussed in
\cite{atwood2}. Some further criticism can be found in
\cite{petrov}.

 Interestingly, the production of
$\eta^\prime$ by thermal gluon fusion at RHIC and the LHC has 
recently been discussed in \cite{jeon} using again the
gluon--$\eta^\prime$ vertex idea. The authors of  \cite{ahmadi} have 
also descibed a gluon fusion process
 to study the inclusive $B\to \eta^\prime X_s$ and the exclusive
 $B\to K^{(*)}\eta^\prime$ decays: the  gluon $g$ of the
 $g g^*\eta^\prime$ vertex is supposed
to be emitted by the light quark inside the $B$ meson, while the
$g^*$ comes from the $b\to s$ penguin. The $B\to K\eta^\prime$
decay has been further investigated in the context of perturbative
QCD in \cite{sanda}.
To assess the reliability of such mechanisms, a study of the gluon
content of the $\eta^\prime$ is in order. The estimate found in
\cite{kou} results in a gluon component as large as $26\%$. This
seems again to favour the possibility that the $\eta^\prime$ can
reliably be coupled through gluons in hadronic processes.

A different scenario proposed to explain the abundant production
of $\eta^\prime$ in $B\to K\eta^\prime$ data was based on the idea
of a strong intrinsic charm component in the $\eta^\prime$
\cite{halperin}. Although  appealing, the $b\to c\bar{c}s$
Cabibbo-favoured process generating an $\eta^\prime$ via its
intrinsic $c\bar{c}$ requires too large a charm component to
explain the data. Recently the charm content of the $\eta^\prime$ has been
thoroughly investigated \cite{poly} and  it is
estimated to be too small to motivate the $b\to c\bar{c}s$ mechanism.

The non-leptonic decays including an $\eta^\prime$ in the final
state, such as $\bar{B}^0 \to D^{0} \eta^\prime$ and 
 $\bar{B}^0\to D^{*0} \eta^\prime$, are also quite interesting  on both 
 theoretical and  experimental grounds.
  Measuring the branching ratios of these
processes may shed light on the nature of the
$\eta$ -- $\eta^\prime$ mixing and of the decay constants $f_\eta$,
$f_{\eta^\prime}$. From the experimental side, at present 
the only known limits on
these decays are  from CLEO \cite{cleo2}.

The simplest possible mechanism  to consider for
 $B \to D \eta^\prime$ decay is
naive  factorization of quark currents as in \cite{rec}. The quark models
considered there do not involve gluons. In contrast, in the
present paper we consider the possibility that a gluon mechanism
could be at work even in a process like $\bar{B}^0\to D^0 \eta^\prime$,
where there is no gluon arising from a penguin diagram
(which is  the case in $B\to K\eta^\prime$). We will assume that
the main contribution to the decay $\bar{B}^0\rightarrow D^0\eta^{'}$
comes from the gluonic mechanism, as illustrated in Fig.~\ref{fig:diag}.
 The $\eta^\prime$ couples to the gluons as shown
in Fig.~\ref{fig:anomaly}.

In general, non-leptonic decays are described by a quark level
effective Lagrangian at some chosen scale. This is built up by
coefficients (containing all short-distance effects above the
chosen scale) times quark operators, typically given as the
product of two currents. In order to get an effective Lagrangian
describing the physical degrees of freedom, the  mesons, a
``bosonization procedure'' is required. This typically amounts to
transforming (see e.g. \cite{hugo}) the original Lagrangian by
extracting collective composite fields with boson quantum numbers.

Historically, bosonizing by factorization of the currents, the 
so-called ``vacuum saturation approximation'' (VSA) has been a
starting point for bosonizing the quark operators. The VSA turns out
to be badly broken in $K$-meson decays. However, in some classes of
heavy meson ($B$) decays, factorization holds up to order $1/m_b$
\cite{BBNS}, but not for the process we consider.

A priori, it would be preferable to keep the description as 
model-independent as possible. Unfortunately, some model-dependence has
to be introduced within our framework. Still the amplitude for $\bar{B}^0
\rightarrow D^0 \eta^\prime$ presented here will be less 
model-dependent than the amplitude for the process
 $B \rightarrow K \eta^\prime$
within the same framework.

One should keep in mind that, when the  four quark operators
responsible for non-leptonic $B$ decays are Fierz-transformed, they
will contain products of coloured currents, such that a virtual
gluon might be emitted from a (light) quark in the $B$ meson due
to such a current.
Furthermore, this virtual gluon might (via the effective
$\eta^\prime g g^*$ vertex) be transformed into an $\eta^\prime$
and a soft gluon. In this paper, this soft gluon is assumed to  make a
gluon condensate with another soft gluon emitted from the light quark
within the $D$ meson.

Inclusion of such gluon condensates gave,
within chiral quark models ($\chi$QM) \cite{chiqm,ERT},
significant effects for $K \rightarrow 2 \pi$ amplitudes, and 
the chiral quark model works well for $K$ decays in 
general \cite{pider,BEF,epb}.
More specific, it has 
been shown that the inclusion of gluon condensates within the
 $\chi$QM works well in order to understand the $\Delta I=1/2$ rule
and the CP-violating quantity $\varepsilon'/\varepsilon$ \cite{BEF}.
(It should also be noted that the value used for the gluon condensate
 in \cite{BEF} is also in fair  agreement with the result obtained
   within generalized factorization \cite{cheng}.)
This gives a strong motivation to use models for mesons containing 
a light and a heavy quark \cite{barhi,effr,itCQM}, and  with soft gluon
 effects in terms of gluon condensates  added  \cite{ahjoe}. This
was already used for the process $D \rightarrow K^0 \bar{K}^0$ in \cite{EFZ}.

\section{Framework for the process $\bar{B}^0\rightarrow D^0\eta'$}

Non-leptonic decays with
$\Delta B=1, \Delta C=\pm 1$
are described by an effective  Lagrangian  at quark level \cite{buras1} :
\begin{equation}
{\cal L}_{eff} = - \fr{G_F}{\sqrt{2}}V_{cb}^*V_{ud}\left[ C_1(\mu)Q_1\,
+ \, C_2(\mu)
Q_2\right]\, ,
\label{Leff}
\end{equation}
where $Q_{1,2}$ are operators given in terms of the quark fields $b,c,u,d$:
\begin{eqnarray}
Q_1&=&\overline{d}\gamma^{\mu}(1-\gamma_5)b\,
\overline{c}\gamma_\mu (1-\gamma_5)u\nonumber \\
Q_2&=&\overline{c}\gamma^{\mu}(1-\gamma_5)b\,
\overline{d}\gamma_\mu (1-\gamma_5)u\, . \label{q1q2}
\end{eqnarray}

The Wilson coefficients $C_1$ and $C_2$ contain  all  
short-distance effects down to the scale $\mu={\cal O}(m_b)$. 
As our quark model calculation of hadronic matrix elements 
will be based on the Heavy Quark Effective Theory 
(HQET), we should in principle
 also calculate the short-distance effects from the scale
${\cal O}(m_b)$ down to a scale ${\cal O}(\Lambda_\chi \sim $ 1 GeV)
and evaluate our matrix elements there \cite{ciuch}
  (also some new operators
 should then appear in the QCD mixing). However, in this work these
complications will be neglected (more on this in section~V).

 The operators defined in eq.~(\ref{q1q2}) may be Fierz-transformed.
 Using the   identity
\begin{equation}
\delta_{\alpha\delta}\delta_{\gamma\beta}=2t^a_{\alpha\beta}
t^a_{\gamma\delta}\, + \, \fr{1}{N_c} \delta_{\alpha\beta}
\delta_{\gamma\delta} \; \; ,
\label{colormatr}
\end{equation}
 the operator
$Q_2$  takes the form :
\begin{figure}[t]
\begin{center}
   \epsfbox{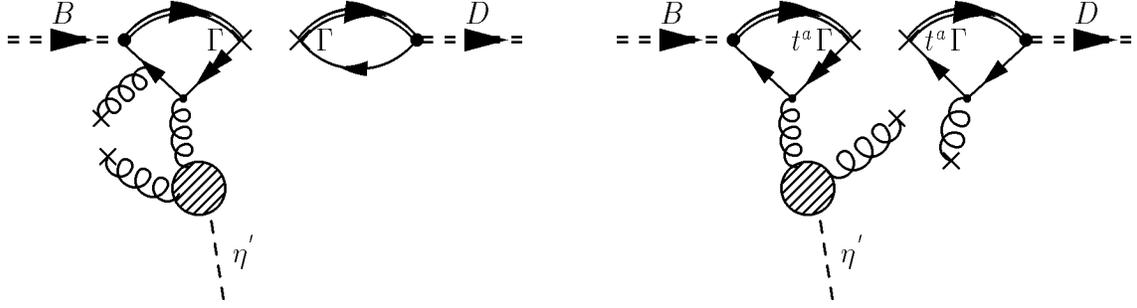}
\epsfysize=14cm \caption{Bosonization of the effective
Hamiltonian, $\Gamma\eq \gamma^\mu (1-\gamma_5)$}
\label{fig:diag}
\end{center}
\end{figure}
\begin{equation}
Q_2=2\overline{c}t^a\gamma^{\mu}(1-\gamma_5)u\,
\overline{d}t^a\gamma_\mu (1-\gamma_5)b\, +\, \fr{1}{N_c}Q_1\, .
\label{q2}
\end{equation}
Fierzing the  operator $Q_1$ gives a similar expression that 
describes just  the process $B\rightarrow D \pi$.

The idea is now that the decay $\bar{B}^0\rightarrow D^0\eta^{'}$ goes in
two steps.  We calculate the amplitude for $\bar{B}^0\rightarrow D^0 g^* g$,
where $g^*$ is the hard virtual gluon and $g$ is a soft gluon.
Then $g^* \rightarrow \eta^\prime g'$, where $g'$ is another soft
gluon. The soft gluons $g$ and $g'$ are then assumed to form a
gluon condensate. For  $\bar{B}^0\rightarrow D^0 g^* g$ we may use a
generalized version \cite{pider,BEF,EFZ} of the VSA,
 and obtain (see eqs.~(\ref{Leff}), (\ref{q1q2}) and  (\ref{q2})):

\begin{eqnarray}
\bra D^0g^{*} g|{\cal L}_{eff}|\bar{B}^0\ket \, =&\, \nonumber \\
\fr{G_F}{\sqrt{2}}&V_{cb}^{*}V_{ud}\left\{
(C_1+\fr{C_2}{N_c})\left[\, \bra D^0 g^{*} g|\overline{c}\gamma_\mu
(1-\gamma_5) u |0\ket\bra 0|\overline{d} \gamma^{\mu}(1-\gamma_5)
b |\bar{B}^0\ket \right.\right. \nonumber \\ &\left.\left.
\qquad\qquad\qquad\qquad +
 \bra D^0|\overline{c}\gamma_\mu (1-\gamma_5)u |0\ket
\bra g^{*} g |\overline{d}\gamma^{\mu}(1-\gamma_5)b|\bar{B}^0\ket\, \right]
\right. \nonumber \\ &\left. \, +
\, 2C_2 \left[\bra g D^0| \overline{c}t^a\gamma_\mu (1-\gamma_5)u | 0\ket
 \bra g^{*} |\overline{d}t^a\gamma^{\mu}(1-\gamma_5)b |\bar{B}^0\ket
\right.\right. \nonumber \\ &\left.\left. \qquad\qquad\qquad +\,
 \bra D^0g^{*}| \overline{c}t^a\gamma_\mu (1-\gamma_5)u | 0\ket
 \bra g |\overline{d}t^a\gamma^{\mu}(1-\gamma_5)b |\bar{B}^0\ket\right]
\right\}\, 
\label{BDggAmp}.
\end{eqnarray}

Within our model, the four terms in (\ref{BDggAmp}) 
  give rise to four  diagrams.
Two of them  are shown in Fig. \ref{fig:diag}, where we have also
included the $\eta^\prime g^* g$ vertex. The other two diagrams are
similar, except for the fact that the hard gluon is emitted from
the $D^0$ meson instead of the $\bar{B}^0$ meson. The first diagram in Fig.
\ref{fig:diag} corresponds to the factorizable part of the
Lagrangian (proportional to $C_1+C_2/N_c$) and  the second
diagram to the non-factorizable part (proportional to $C_2$).
The dominant contribution comes from the non-factorizable part
 because of the large  ratio of the
Wilson coefficients $2C_2/(C_1 + C_2/N_ c) \simeq 12$
at $\mu = m_b$. It should also be noted that $C_2$ is rather stable with
respect to variations in the renormalization scale $\mu$.

\begin{figure}[t]
\begin{center}
   \epsfbox{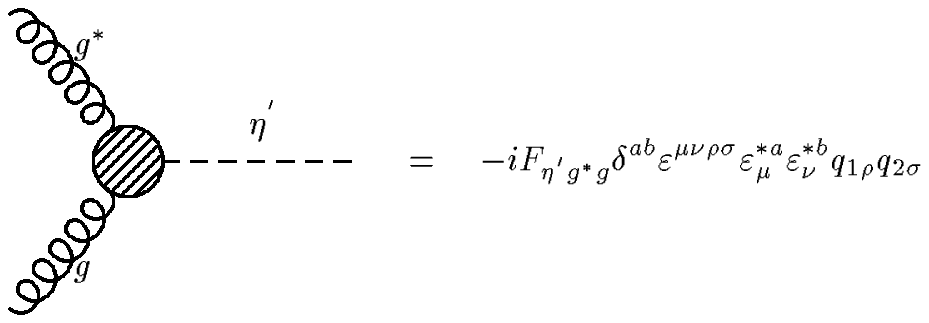}
\epsfysize=10cm
\caption{$\eta^{'}g^*g$\, -\, vertex}
\label{fig:anomaly}
\end{center}
\end{figure}
The form factor$F_{\eta^{'}g^{*}g}$ has been examined by many
groups following \cite{atwood,hou} (see also \cite{molti}), and takes
the form given in Fig.~\ref{fig:anomaly} :

In the Fock--Schwinger gauge one  replaces
$\varepsilon_\nu^{*b}\rightarrow
-\fr{i}{2}\fr{\partial}{\partial q_{1\rho}} G_{\rho\nu}^{b}(0)$ \cite{nov},
so that the $\eta^{'}g^*g$  vertex now takes the form:

\begin{equation}\label{etagg}
 -\fr{1}{2}F_{\eta^{'}g^{*}g} \; \delta^{ab}
\varepsilon^{\mu\nu\rho\sigma}\varepsilon_\mu^{*a}G_{\nu\rho}
^b(0)q_{\sigma}\; ,
\end{equation}
where $q$ is the momentum of the virtual gluon.
We will use the parametrization of Hou and Tseng \cite{hou}
 where the form factor is :
\begin{equation}
F_{\eta^{'}g^{*}g}=\sqrt{\fr{3}{2}}\fr{\alpha_s(q^2)}{\pi
f_\pi}\, .
\end{equation}

\section{Elements of the Heavy--Light Chiral Quark Model (HL$\chi$QM)}

Some of the  matrix elements in eq. (\ref{BDggAmp}) are known
(e.g. the ones given by the decay constants $f_D$ and $f_B$), and
some are model-dependent within our approach. Our calculation will,
to some extent, be based on what we will call a Heavy Light Chiral
Quark Model (HL$\chi$QM) \cite{ahjoe} (see also \cite{EFZ}).
 This is a quark loop model
supposed to describe interactions between quarks with momenta
below the chiral symmetry breaking scale $\Lambda_\chi \sim$ 1
GeV. The model is similar
to the ones that  have
been used by several groups \cite{barhi,effr,itCQM}, but it contains in
 addition soft gluons which may form gluon condensates, similar to one
 version of the chiral quark model in the light sector \cite{pider,BEF,ERT}.
The key ingredient in this kind of models is a phenomenological
term in the Lagrangian, which couples the quark fields directly to
the meson fields. This enables one to integrate out the quark
degrees of freedom and obtain a Lagrangian containing only meson
fields \cite{ebert1}. A drawback with these models is that they do not
incorporate any confining mechanism for quarks. However, such
models, as the related Nambu-Jona-Lasinio
 models \cite{barhi,effr,itCQM,bijnens},
describe chiral symmetry breaking reasonably well, and that is
important for our purpose. As for the light sector
\cite{pider,BEF,chiqm,bijnens}, the coefficients of various pieces of
 chiral Lagrangians can be calculated.

The Lagrangian of the HL$\chi$QM is
\be
{\cal L}={\cal L}_{HQET} + {\cal L}_{\chi QM} +
{\cal L}_{Int} + ... \; ,
\label{totlag}
\ee
where
\begin{equation}
{\cal L}_{HQET}=\overline{Q_v} \, i v \cdot D \, Q_v + {\cal
O}(m_Q^{-1})
 \label{LHQET}
\end{equation}
is the Lagrangian for Heavy Quark Effective Field Theory (HQEFT).
The heavy quark field  $Q_v$
annihilates a heavy quark with velocity $v$ and mass $m_Q$, and $D_\mu$ is
a covariant derivative containing the gluon field and the photon field.

In the pure light sector, the Lagrangian of the chiral quark model
($\chi$QM) can be written as :
\be
{\cal L}_{\chi QM}= \chibar \left[\gamma^\mu (i D_\mu \, + \,
{\cal V}_{\mu} \,+\, \gamma_5  {\cal A}_{\mu}) \,  - \, m
\right]\chi \, .
 \label{chqmR}
 \ee
 where $m$ is the constituent
mass (200 -- 300 MeV) of the light quarks. The field $\chi$ transforms
under the unbroken $SU(3)_V$ symmetry, so that the quark fields
transforming under $SU(3)_L \times SU(3)_R$ are
\begin{equation}\label{rotated}
q_L=\xi^\dagger\chi_L \;  \qquad \text{and} \qquad q_R=\xi\chi_R\, ,
\end{equation}
where $\xi$ is a $3\times 3$ matrix containing the Nambu--Goldstone
bosons of QCD ($\pi, K, \eta$):
\begin{equation}
\xi=e^{-i\Pi/f}\;\, ,  \quad
 \text{where}\, \quad \Pi=\fr{\lambda^a}{2}\phi^a(x) = \frac{1}{\sqrt{2}}
\left[\begin{array}{ccc} \fr{\pi^0}{\sqrt{2}}+\fr{\eta_8}{\sqrt{6}} & \pi^+
&K^+\\ \pi^-&-\fr{\pi^0}{\sqrt{2}}+\fr{\eta_8}{\sqrt{6}} & K^0\\
K^- &\overline{K^0}& -\fr{2}{\sqrt{6}}\eta_8\end{array}\right] \; .
\end{equation}
Moreover,
 $f\simeq f_\pi\,=\, 93.2$ MeV
 and $\lambda^a, \, a=1\ldots 8$ are
the Gell-Mann matrices.
 The vector and axial vector fields are
\begin{equation}
{\cal V}_{\mu}\eq \fr{i}{2}(\xi\partial_\mu\xi^\dagger
+\xi^\dagger\partial_\mu\xi)
\;  \qquad \text{and} \qquad
{\cal A}_\mu \eq \fr{i}{2}
(\xi\partial_\mu\xi^\dagger -\xi^\dagger\partial_\mu\xi) \; .
\label{defVA}
\end{equation}

The meson--quark interaction for the heavy--light system is
\cite{barhi,effr,itCQM,ahjoe}:
\be
{\cal L}_{Int} =
- G_H \, \left[ \chibar_f \, \overline{H_{vf}} \, Q_v \,
 +  \, \overline{Q_v} \, H_{vf} \, \chi_f \right] \, ,
\label{Int}
\ee
where $G_H$ is a
 coupling constant (in \cite{barhi,effr,itCQM} it corresponds to a
renormalization constant), and
 $H_{f}$ is a field containing both the singlet ($0^-$)
and triplet ($1^-$) meson field:

\begin{eqnarray}
&H_{f} \eq P_{+} \left( P_{\mu f} \gamma^\mu - i P_{5f} \gamma_5
\right) \nonumber \\ &\overline{H_{f}}\eq\gamma^0 (H_{f})^\dagger
\gamma^0 = \left( (P_{\mu f})^{\dagger} \gamma^\mu - i
(P_{5f})^\dagger \gamma_5 \right) P_+ \; , \label{barH}
\end{eqnarray}
where
\begin{equation}
P_+=(1 + \slash{v})/2 
\label{proj}
\end{equation}
is a projection operator. The index $f$ runs over the light quark
flavours $u, d, s$. The field $P_{5}(P^\mu)$ annihilates a heavy meson
with spin-parity $0^-(1^-)$, with velocity $v$.

The way of regularizing divergent loop integrals has to be taken
as a part of the definition of the model. One might use some kind
of cut-off regularization (sharp cut-off as in \cite{barhi} and 
cut-off in the proper time formalism as in \cite{effr,itCQM}), cutting  the
momenta in the loop between $\Lambda_\chi\sim 1$ GeV and some
infrared cut-off $\mu_{IR}$. The infrared cut-off could be zero as
in \cite{effr} or $\sim m$ as in \cite{barhi,itCQM}. Still,
dimensional regularization may be used if there are no mass scales
bigger than $\Lambda_\chi$ entering the loop.
Within our model we will relate all divergent loop integrals to some
physical parameter (as for instance $f_\pi$), as was done
 in \cite{pider,BEF,ERT}.
 This means in particular that we will treat
quadratic, linear and logarithmic  divergences as different.
 The type of divergence can easily be seen by
inspection.

The calculation of soft gluon effects in terms of gluon condensates
is carried out in the
Fock--Schwinger gauge. In this gauge one can expand the gluon field as :
\be
A_\mu^a(k)=-\fr{i(2\pi)^4}{2}G_{\rho\mu}(0)\fr{\partial}{\partial
k_\rho}\delta^{(4)}(k) + \cdots\;\, , 
\ee
where $k$ is the gluon momentum which is put to zero after
the derivation is performed.
The fact that the soft gluon tensor is appearing explicitly in this
expression makes it a powerful tool to derive gluon condensate
contributions to many processes, as  demonstrated in \cite{nov}.
\begin{figure}[t]
\begin{center}
\epsfbox{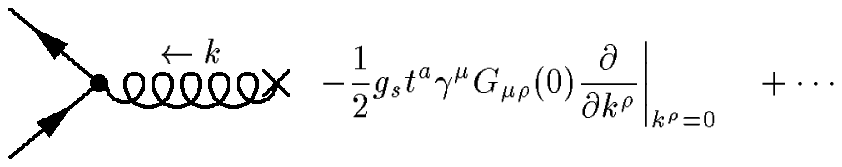}
\epsfysize=5cm
\caption{Feynman rule for a quark coupling to a soft gluon.}
\label{fig:gc}
\end{center}
\end{figure}
Since each vertex in a
Feynman diagram is followed  by an integration over loop momentum,
 we get the
Feynman rule for interactions between light quarks and soft gluons
shown in Fig.~\ref{fig:gc}.  From
this Feynman rule it immediately follows that the coupling
of a soft gluon
 to a heavy quark is suppressed by $1/m_Q$, since the
vertex to this order  is proportional to $v_\mu v_\nu
G^{a\mu\nu}\,=\, 0$, $v_\mu$ being the heavy quark velocity.

\begin{figure}[t]
\begin{center}
\epsfbox{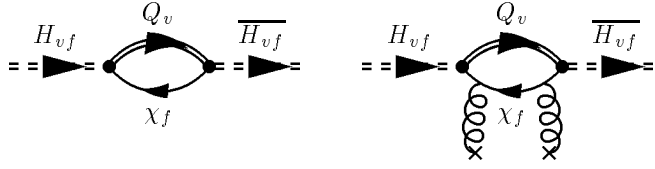}
\epsfysize=10cm
\caption{Self-energy for a heavy--light meson}
\label{fig:self}
\end{center}
\end{figure}

The diagrams contributing to the kinetic term of heavy--light
mesons are shown in Fig.~\ref{fig:self}. By including also diagrams
with the insertion of the $\cal{A}_\mu ({\cal V}_\mu )$ fields on
light quark loop lines, one obtains the low energy effective
Lagrangian for heavy mesons coupling to the light mesons  $\pi, K,
\eta, \ldots$ in the usual form\cite{wise1,itchpt} :
\begin{equation}
{\cal L}_{\mbox{Str}}\, = \, 
- \mbox{Tr}\left[\overline{H_{a}}(iv\cdot D -\Delta
)H_{a}\right]\, +\, \mbox{Tr}\left[\overline{H_{a}}H_{b}v^\mu {\cal
 V}_{\mu ba}\right] \, 
- \, g_A \mbox{Tr}\left[\overline{H_{a}}H_{b}\gamma^\mu\gamma_5
{\cal A}_{\mu ba}\right]\; ,
 \label{LStr}
\end{equation}
where $\Delta$ is the mass difference between the heavy meson(s) and 
the  corresponding heavy quark. 
The trace is only taken over the gamma matrix indices. Note that
the ${\cal V}_\mu$ field and the covariant derivative $D_\mu$
 (containing the photon field) can be combined into
a total covariant derivative
 ${\cal D}_{\mu ba}\eq D_\mu\delta_{ba} + i {\cal V}_{ba}$.
 The heavy--light meson fields, $P_{5f}$ and $P_{\mu f}$,
 have been rescaled by
a factor of $\sqrt{M_P}$ and $\sqrt{M_{P^*}}$. The mass dimension
of the heavy--light meson fields is now 3/2. Typical values of 
the axial coupling $g_A$ found in the literature lie between 1/4 and 2/3.

In order to obtain eq. (\ref{LStr}), we have to identify 
the square of the coupling $G_H$ in (\ref{Int}), multiplied by some
 loop integrals with the coefficients $g_V \equiv 1$ for the kinetic
term (including the vector term) and $g_A$ for the axial vector term.
Such loop integrals
 contain logarithmic and linear divergences plus
finite terms (gluon condensate terms are
 finite in this case). Details
will be given in \cite{ahjoe}.
The logarithmic divergent integral appearing here is the same as the
one appearing in the pure light sector \cite{pider,BEF,ERT}, and is related
to $f_\pi$.
The quantity $\delta g_A \equiv (1 - g_A)$ is related to
a linear divergent integral plus finite terms (including  gluon condensates).
 Eliminating the linearly 
 divergent integral (appearing as finite in dimensional regularization)
and identifying the logarithmic divergent integral with $f_\pi$ as
in the pure light case, we obtain  a relation for $G_H$ in terms
of $f_\pi$, the constituent light quark mass $m$, $\delta g_A$,
and the gluon condensate \cite{ahjoe} :

\begin{equation}\label{relgh}
G_H^2=\fr{2 m(1 - \frac{3}{4} \delta g_A)}
{f_\pi^2+\fr{N_cm^2}{8\pi}-\fr{\kappa}{2m^2}
\bra\fr{\alpha_s}{\pi} G^2\ket} \;\, , \qquad \text{where} \qquad
\kappa\eq\fr{96 - 3\pi}{768} \; .
\end{equation}
Thus, in the formal limit where only the logarithmic divergent
integral is kept, one obtains the  simple  expression \cite{EFZ}:
\begin{equation}\label{relgh-simp}
G_H \simeq \fr{\sqrt{2 m}}{f_\pi} \; .
\end{equation}

 The
gluon condensate and the constituent quark mass can be linked to
the mass difference between the $0^-$ and $1^-$ states via the
matrix element of the colour magnetic operator 
between  heavy states \cite{MagnDip}.
 Calculating this matrix element in our model
\cite{ahjoe}  gives a   gluon condensate range (301 MeV)$^4 <
\bra \fr{\alpha_s}{\pi} G^2\ket < $ (327 MeV$)^4$ 
for a constituent light quark mass between
$230$ and $270$ MeV. Then we find that
also numerically, the simple expression (\ref{relgh-simp})
 is close to eq.~(\ref{relgh})
because the second term ($\sim N_c m^2$) and third term 
($\sim \bra\fr{\alpha_s}{\pi} G^2\ket$) in the denominator of
eq.~(\ref{relgh}) tend to cancel.

 When calculating the  diagram  for $\bar{B}^0 \rightarrow g^*$,
the virtual gluon is not soft. (The light quark emerging from the weak
vertex carries a momentum of order $2$ GeV.) In order to avoid
mass scales bigger than $\Lambda_\chi$ in the loop integral, as
explained earlier, one has
 to split off this high momentum
in terms of a Large Energy Effective Theory (LEET)
\cite{charles,aglietti,bauer} for the quark propagating between
the weak vertex and the $\eta^\prime$ vertex (indicated by two
arrows in Fig.~\ref{fig:diag}).
 This is  analogous to  splitting off  the heavy
quark mass  within the HQET. The momentum of the quark is written  as
$p = E_{\eta^\prime} n + k$,
 where $E_{\eta^\prime}$ is the energy
of the $\eta^\prime$ and $n^\mu=(1;0,0,1)$ is parallel to the four
momenta of the $\eta^\prime$. Furthermore, $k$ is the residual
momentum which has to be
 less than $\Lambda_\chi \sim$ 1 GeV.

The LEET  tells us that the effect of a soft gluon that  couples
 to a fast
light quark is suppressed by $1/E_{\eta^\prime}$, since the
leading order coupling  is proportional to $n_\mu n_\nu
G^{a\mu\nu}\,=\,0$.
 Neglecting $1/m_Q$ and $1/E_{\eta^{'}}$ corrections, soft
gluons can only couple to the spectator quark in the $B$ and in the
$D$ meson.

\section{The Amplitude for $\bar{B}^0 \rightarrow D^0 \eta^\prime$}

The first part of the diagram to the right in Fig.~\ref{fig:diag}
describes the  matrix element for the transition
 $\bar{B}^0\rightarrow g^*$, where the gluon is off shell and
carries a momentum $q$, with $q^2\,=\, m_{\eta^\prime}^2$.
Within our model, light vector mesons $V=(\rho,\omega,\phi)$ might couple
to light quarks like the field ${\cal V}_{\mu}$ in eq.~(\ref{chqmR})
by letting ${\cal V}_{\mu} \to ({\cal V}_{\mu} - \rho_\mu)$ \cite{itchpt}.
Thus, since the gluon
is a vector particle, the transition $\bar{B}^0\rightarrow g^*$
is, within our model,
related to the transition $\bar{B}^0\rightarrow V=(\rho,\omega,\phi)$.
In other words, writing down  the loop integral
for $\bar{B}^0\rightarrow g^*$ and for $\bar{B}^0\rightarrow V$,
 the  integrals are the same,
 only the couplings are different. Therefore we can avoid going into 
details about the loop integral because the relevant form factor
for $\bar{B}^0\rightarrow \rho$ is known from 
lattice and light cone sum rules \cite{ukqcd}.
Using this fact will reduce the model dependence of our calculation.
 Considering a vector meson dominance (VMD)
induced coupling of the $\rho$ to the light quarks, $\gamma^\mu
m_\rho^2 /f_\rho$ (see \cite{polosa1} for more details), we have
in the LEET approximation:
\begin{eqnarray}
J^{\mu a}(\bar{B}^0\rightarrow g^{*d}) \eq & \bra g^{*d}
|\bar{d}t^a\gamma^{\mu}(1-\gamma_5)b |\bar{B}^0\ket\nonumber \\
\simeq & \fr{\delta^{ad}}{2} \fr{g_s f_\rho}{m_\rho^2}\bra \rho
|\bar{d}\gamma^{\mu}(1-\gamma_5)b |\bar{B}^0\ket\, ,
\label{rhocurrent}
\end{eqnarray}
where $m_\rho$ is the $\rho$ mass and $f_\rho \,=\, 0.152$ GeV$^2$.
 Furthermore, $a$ and $d$ are  colour indices from the
 quark-gluon vertex and the factor $\delta^{ad}/2$ comes from the trace
over the $SU(3)_c$ matrices. This current can now be parametrized
by the form factors from the LEET  result for $\bar{B}^0\rightarrow
\rho$ \cite{charles} :
\begin{equation}
J^{\mu a}(\bar{B}^0\rightarrow g^{*d})
 = \fr{\delta^{ad}}{2}  \fr{g_sf_\rho}{m_\rho^2}
2E_{\eta^\prime}\zeta_{\perp}
\left[i \varepsilon^{\mu\nu\rho\sigma} v_\nu n_\rho
\varepsilon^{*}_\sigma - (n \cdot v)
\varepsilon^{*\mu}+(\varepsilon^* \cdot v)n^\mu\right]\, .
\label{nonfact}
\end{equation}

A relation corresponding to (\ref{rhocurrent}) cannot be used for
the process $B \rightarrow K \eta^\prime$.
 The other part of the diagram, with
a light quark moving in the background field of soft gluons,
contains a logarithmic divergence, which can be related to $f_\pi$,
the result of the calculation is \cite{ahjoe,EFZ}:
\bea
\bra g D^0| \overline{c}t^a\gamma_\mu (1-\gamma_5)u | 0\ket \, &=& \,
-\fr{G_Hg_s\sqrt{M_D}}{8m^2N_c}G^{a}_{\gamma\delta}\left[
\fr{m^2N_c}{4\pi}
(v'^{\delta}g^{\mu\gamma}-v'^{\gamma}g^{\mu\delta})\nonumber\right. \\ &&\left.
-i\varepsilon^{\mu\alpha\gamma\delta}v'_\alpha
\left(f_\pi^2+\fr{m^2N_c}{4\pi}\right)\right]\, ,\label{nonfact2}
\eea
where $v'$ is the velocity of the $D$ meson. Combining eqs.~(\ref{etagg}),
 (\ref{nonfact}) and (\ref{nonfact2}), we find for
the non-factorizable contribution to $\bar{B}^0\rightarrow D^0\eta^{'}$:
\begin{eqnarray}
\label{nfact}
 A^{NF} \eq C_2 \, \pi \, G_H \, \sqrt{M_D} \, 
\frac{E_{\eta^\prime}^2   F_{\eta^\prime g^*g} \, \zeta_\perp
 f_\rho}{6 m_{\eta^\prime}^2 m_\rho^2} (n \cdot v) (n \cdot v')
\left( 1 + \frac{n\cdot v}{n \cdot v'} \right)
\left(\fr{2 \pi f_\pi^2}{m^2N_c} + 1 \right)
\bra \fr{\alpha_s}{\pi}G^2 \ket\, .
\end{eqnarray}
In the calculation we have used $q^2=m_{\eta^\prime}^2$ in the
gluon propagator. In the parenthesis  $(1 + n\cdot v/n \cdot v')$
above, the term $1$ is due to the diagram to the right in Fig.~\ref{fig:diag},
and the term  $n\cdot v/n \cdot v$ is due to the corresponding diagram
where the $\eta^\prime$ couples to the
$D$ meson. The inclusion of this diagram
deserves a comment. The form factor $\zeta_\perp$ for the transition
 $\bar{B}^0\rightarrow \rho$ contains a factor $\sqrt{M_B}$ \cite{charles},
and it will also contain a factor $1/(n \cdot v)$ from the LEET
loop integral. In this case there is also a factor $\sqrt{M_D}$
coming from soft gluon emission from the light quark in the
$D$ meson (see eq.~(\ref{nonfact2})). For the second diagram we have
to interchange $M_B \leftrightarrow M_D$, and $v \leftrightarrow
v'$. In total we obtain (\ref{nfact}), where $\zeta_\perp$ is the
$\bar{B}^0\rightarrow \rho$ transition form factor.

The factorizable contributions are described by the diagram to the
left in Fig.~\ref{fig:diag}. The $B \rightarrow D g g^*$
 part of the diagram (connected to
the $\eta^\prime$ vertex) is finite. The second part is the
current $\bra D^0|\overline{c}\gamma_\mu(1-\gamma_5)u |0\ket$, which
is  by definition given by the decay constant
 $f_D$. The finite diagrams have to be calculated using the
LEET,  and the result of the calculation is:
\begin{eqnarray}\label{fact}
A^F \eq - \left(C_1 +\fr{C_2}{N_c}\right) \, \pi \, G_H \, 
\fr{E_{\eta^\prime}F_{\eta^\prime g^*g}}{288 \, m \, m_{\eta^\prime}^2}
\left[ f_D\sqrt{M_D} \, n\cdot v'+f_B\sqrt{M_B} \, n \cdot v\right]
\bra  \fr{\alpha_s}{\pi}G^2 \ket\, .
\end{eqnarray}

\section{Numerical results}

The total amplitude due to the gluon mechanism considered in this paper is:
\begin{equation}\label{totamp}
\bra D^0\eta^{'}| {\cal L}_{eff}|\bar{B}^0\ket\, =\,
- \fr{G_F}{\sqrt{2}}V_{cb}^{*}V_{ud}( A^{F}\, +\, A^{NF})\, ,
\end{equation}
where $A^F$ and $A^{NF}$ are given in eqs. (\ref{fact}) and
(\ref{nfact}). In the LEET limit the kinematics is easy:
\begin{eqnarray}
n\cdot v\, =\, 1 \; , \qquad n\cdot v' \, =\, M_B/M_D \; , \qquad
E_{\eta^{'}}\, =\, (M_B^2-M_D^2)/(2M_B)\, .
\end{eqnarray}
We will use $M_B =$ 5.280 GeV, $M_D =$ 1.870 GeV,
$m_{\eta^\prime} = $ 0.958 GeV , $G_F = 1.166 \times
 10^{-5}$ GeV$^{-2}$, $~V_{cb}^*V_{ud} = 0.037$ and $f_\pi = $ 93.2 MeV,
which can be found in PDG \cite{pdg}. The value of $\alpha_s$ has
to be taken at the point $q^2=m_{\eta^\prime}^2$, where
 $\alpha ({m_{\eta^\prime}}^2)\simeq
0.56$\cite{www}. The Wilson coefficients, $C_1$ and $C_2$, have
been calculated \cite{buras1} in NLO to be
$C_1(m_b)\, =\, -0.184$ and $C_2(m_b)\,=\,1.078$
(where the  NDR scheme and  $\Lambda^{(5)}_{\overline{MS}}\, =\,$ 225
MeV have been used).

Extrapolating $C_{1,2}$ naively down to $\mu\e\Lambda_\chi$,
we find that $C_2$ does not change much,
$C_2(\Lambda_\chi)  \e 1.1695$, while
 $C_1$ is more sensitive,
$C_1 (\Lambda_\chi )\e -0.3695$. This gives
 $2C_2/(C_1 \p C_2/N_c) \, \simeq\, 98$ at $\Lambda_\chi$.
However, this number should not be taken too seriously. Using HQEFT between
the scales $m_b$ and $\Lambda_\chi$, there are more operators
entering as in \cite{ciuch}.
 However, the important thing for our calculation is that the coefficient
$C_2$ of the dominating operator $Q_2$ is rather stable with respect to
 variations of the scale.

We extract the numerical value for the LEET form factor $\zeta_\perp$
from 
 lattice and light-cone sum-rule results \cite{ukqcd}, from
which we obtain the  value $\zeta_\perp \, \simeq \, 0.32.$

The only model-dependent parameters entering  eq.~(\ref{totamp}) are
the gluon condensate and the constituent light quark mass $m$.
As mentioned above,  the range
$230 < m < 270$ MeV for the constituent quark mass gives the range
 (301 MeV)$^4 < \bra \frac{\alpha_s}{\pi} G^2\ket < $ (327 MeV$)^4$  
for the gluon condensate.

Inserting all these values in (\ref{totamp}), and including the uncertainty in
$C_2$, we obtain the  decay rate :
\begin{equation}
\Gamma (\bar{B}^0\rightarrow D^0\eta^{'}) \,=\,
\fr{M_B^2-M_D^2}{16\pi M_B^3}|\bra D^0\eta^{'}| {\cal L}_{eff}|\bar{B}^0\ket|^2\,=\,
(0.7 \, \mbox{--} \, 1.4) \times 10^{-16} \text{GeV} \, ,
\label{rate}
\end{equation}

which gives a branching ratio
\begin{equation}
Br(\bar{B}^0\rightarrow D^0\eta^{'}) \,= 
\, (1.7 \, \mbox{--} \, 3.3) \times 10^{-4}.
\label{branch}
\end{equation}

 The ranges in (\ref{rate}) and (\ref{branch}) come from the variation of the
 constituent light quark mass
$m$ in the range 230 -- 270 MeV, and the uncertainty in $C_2$,
 i.e. the uncertainty within
the model. In addition, there is  the  uncertainty from applying this
model.

\section{Conclusions}

In this paper we have computed the $\bar{B}^0\to D^0\eta^\prime$ exclusive
process coupling the $\eta^\prime$ via its gluonic component and
exploiting the idea of a glue--glue--$\eta^\prime$ effective vertex,
 connected to the gluonic triangle anomaly.
The calculation is performed using a heavy--light chiral quark
model containing soft gluons which may form gluon condensates. Such a
model is an extension of constituent quark models of the type 
 obtained by bosonizing an underlying Nambu--Jona--Lasinio
interaction involving heavy and light quarks.

The results found, when compared with those in \cite{rec}, where a
branching ratio 
$Br(\bar{B}^0~\rightarrow~D^0\eta') \,\simeq\, 0.30\times10^{-4}$ is
obtained, show remarkably that the gluon mechanism could be more
important than the one obtained from standard  factorization of quark
 currents.  It would be very interesting
to check this approach directly on new data on exclusive $B$
decays to $\eta^\prime$ mesons. At the moment  only an upper bound is known 
 for this exclusive branching ratio, amounting to
$Br(\bar{B}^0\rightarrow D^0\eta^{'}) \,< \, 9.4\times10^{-4}$ \cite{pdg}.

The relevance of the gluonic cloud in the $\eta^\prime$, here
considered from a phenomenological standpoint, is a crucial
question within the theory. The discrepancy in
mass between the $\eta^\prime$ state and the octet of
($\pi,K,\eta$) goldstones can directly be attributed to the gluonic content of
$\eta^\prime$ \cite{georgi}. Moreover, the Witten--Veneziano mass
formula for the $\eta^\prime$ \cite{wv}, elaborated in the $1/N$
language, has  very recently been derived on the lattice
\cite{giusti} confirming in this set up the profound relation
between $m_{\eta^\prime}$ and the breaking of the $U(1)_A$
symmetry due to the gluon anomaly.

\vspace{1cm}
ADP acknowledges support from the EU-TMR programme, contract
CT98-0169. AH acknowledges the hospitality at the CERN TH Division
where this work was completed. JOE thanks P. Ball for useful comments on
ref. \cite{ukqcd}.

\bibliographystyle{unsrt}

\end{document}